\documentclass[useAMS,usegraphicx,usenatbib]{mn2e}                
\usepackage{times}

\newcommand{\Msun}{\ensuremath{\,{\rm M}_\odot}}                  
\newcommand{\Rsun}{\ensuremath{\,{\rm R}_\odot}}                  
\newcommand{\Teff}{\ensuremath{T_{\rm eff}}}                      
\newcommand{\Mjup}{\ensuremath{\,{\rm M}_{\rm Jup}}}              
\newcommand{\Rjup}{\ensuremath{\,{\rm R}_{\rm Jup}}}              
\newcommand{\Teq}{\ensuremath{T_{\rm eq}^{\,\prime}}}             
\newcommand{\safronov}{\ensuremath{\Theta}}                       
\newcommand{\kms}{\,km\,s$^{-1}$}                                 
\newcommand{\ms}{\,m\,s$^{-1}$}                                   
\newcommand{\mss}{\,m\,s$^{-2}$}                                  
\newcommand{\as}{\ensuremath{^{\prime\prime}}}                    
\newcommand{\am}{\ensuremath{^\prime}}                            
\newcommand{\FeH}{\ensuremath{\left[\frac{\rm Fe}{\rm H}\right]}} 
\newcommand{\pjup}{\ensuremath{\,\rho_{\rm Jup}}}                 
\newcommand{\psun}{\ensuremath{\,\rho_\odot}}                     

\newcommand{\mcc}[1]{\multicolumn{3}{c}{#1}}

\newcommand{\ermcc}[5]{\mcc{\ensuremath{{#1\,^{+#2}_{-#3}}\,^{+#4}_{-#5}}}}

\newcommand{\prism}{\textsc{prism}}	
\newcommand{\gemc}{\textsc{gemc}}	

\title[Transits and starspots of WASP-6]
{Transits and starspots in the WASP-6 planetary system}

\author[Tregloan-Reed et al.]
       {Jeremy Tregloan-Reed\,$^{1,2}$\thanks{Email: jeremy.j.tregloan-reed@nasa.gov},
        John Southworth\,$^{2}$, 
        M.\ Burgdorf\,$^{3}$,                        
        S.\ Calchi Novati\,$^{4,5,6}$,       
        \newauthor
        M.\ Dominik\,$^{7}$\thanks{Royal Society University Research Fellow},
        F.\ Finet\,$^{8,9}$,                      
        U.\ G.\ J{\o}rgensen\,$^{10}$, 
        G.\ Maier\,$^{11}$,                            
        L.\ Mancini\,$^{12}$,                              
        S.\ Proft\,$^{11}$,                                
        \newauthor
        D.\ Ricci\,$^{13,14,15}$,                              
        C.\ Snodgrass\,$^{16}$,                           
        V.\ Bozza\,$^{5,17}$, 
        P.\ Browne\,$^{7}$, 
        P.\ Dodds\,$^{7}$, 
        T.\ Gerner\,$^{11}$, 
        \newauthor
        K.\ Harps{\o}e\,$^{10}$, 
        T.\ C.\ Hinse\,$^{18,19}$, 
        M.\ Hundertmark\,$^{7}$, 
        N.\ Kains\,$^{20}$, 
        E.\ Kerins\,$^{21}$, 
        C.\ Liebig\,$^{7}$, 
        \newauthor               
        M.\ T.\ Penny\,$^{22}$, 
        S.\ Rahvar\,$^{23}$, 
        K.\ Sahu\,$^{20}$, 
        G.\ Scarpetta\,$^{6,5,17}$, 
        S.\ Sch\"afer\,$^{24}$, 
        F.\ Sch\"onebeck\,$^{11}$, 
        \newauthor                
        J.\ Skottfelt\,$^{10}$, 
        J.\ Surdej\,$^{8}$, 
        \\
        \\
        $^{1}$\,NASA Ames Research Center, Moffett Field, CA 94035, USA \\
        $^{2}$\,Astrophysics Group, Keele University, Staffordshire, ST5 5BG, UK \\
        $^{3}$\,Universit\"at Hamburg Meteorologisches Institut Bundesstra{\ss}e 55 20146 Hamburg, Germany \\
        $^{4}$\,NASA Exoplanet Science Institute, MS 100-22, California Institute of Technology, Pasadena, CA 91125, US \\
        $^{5}$\,Dipartimento di Fisica ``E.R. Caianiello'', Universit\`a di Salerno, Via Giovanni Paolo II 132, 84084, Fisciano (SA), Italy \\
        $^{6}$\,Istituto Internazionale per gli Alti Studi Scientifici (IIASS), 84019 Vietri Sul Mare (SA), Italy \\
        $^{7}$\,SUPA, University of St Andrews, School of Physics \& Astronomy, North Haugh, St Andrews, KY16 9SS, UK \\
        $^{8}$\,Institut d'Astrophysique et de G\'eophysique, Universit\'e de Li\`ege, 4000 Li\`ege, Belgium \\
        $^{9}$\,Aryabhatta Research Institute of Observational Sciences (ARIES), Manora Peak, Nainital-263 129, Uttarakhand, India \\
        $^{10}$\,Niels Bohr Institute \& Centre for Star and Planet Formation, University of Copenhagen, \O stervoldgade 5, 1350 Copenhagen K, Denmark \\
        $^{11}$\,Astronomisches Rechen-Institut, Zentrum f\"ur Astronomie, Universit\"at Heidelberg, M\"onchhofstra{\ss}e 12-14, 69120 Heidelberg, Germany \\
        $^{12}$\,Max Planck Institute for Astronomy, K\"onigstuhl 17, 69117 Heidelberg, Germany \\
        $^{13}$\,Observatorio Astron\'omico Nacional, Instituto de Astronom\'ia -- Universidad Nacional Aut\'onoma de M\'exico, Ap. P. 877, Ensenada, BC 22860, Mexico \\
        $^{14}$\,Instituto de Astrof\'isica de Canarias, E-38205 La Laguna, Tenerife, Spain \\
        $^{15}$\,Universidad de La Laguna, Departmento de Astrof\'isica, E-38206 La Laguna, Tenerife, Spain \\
        $^{16}$\,Planetary and Space Sciences, Department of Physical Sciences, The Open University, Milton Keynes, MK7 6AA, UK \\
        $^{17}$\,Istituto Nazionale di Fisica Nucleare, Sezione di Napoli, 80126 Napoli, Italy \\
        $^{18}$\,Korea Astronomy and Space Science Institute, Daejeon 305-348, Republic of Korea \\
        $^{19}$\,Armagh Observatory, College Hill, Armagh BT61 9DG, UK \\
        $^{20}$\,Space Telescope Science Institute, 3700 San Martin Drive, Baltimore, MD 21218, USA \\
        $^{21}$\,Jodrell Bank Centre for Astrophysics, University of Manchester, Oxford Road, Manchester M13 9PL, UK \\
        $^{22}$\,Department of Astronomy, Ohio State University, 140 W. 18th Ave., Columbus, OH 43210, USA \\
        $^{23}$\,Department of Physics, Sharif University of Technology, P.\,O.\,Box 11155-9161 Tehran, Iran \\
        $^{24}$\,Institut f\"ur Astrophysik, Georg-August-Universit\"at G\"ottingen, Friedrich-Hund-Platz 1, 37077 G\"ottingen, Germany \\
        \vspace*{-0.8cm}
}

\begin{document} \maketitle 

\begin{abstract}
We present updates to \prism, a photometric transit-starspot model, and \gemc, a hybrid optimisation code combining MCMC and a genetic algorithm. We then present high-precision photometry of four transits in the WASP-6 planetary system, two of which contain a starspot anomaly. All four transits were modelled using \prism\ and \gemc, and the physical properties of the system calculated. We find the mass and radius of the host star to be $0.836\pm 0.063\Msun$ and $0.864\pm0.024\Rsun$, respectively. For the planet we find a mass of $0.485\pm 0.027\Mjup$, a radius of $1.230\pm0.035\Rjup$ and a density of $0.244\pm0.014\pjup$. These values are consistent with those found in the literature. In the likely hypothesis that the two spot anomalies are caused by the same starspot or starspot complex, we measure the star’s rotation period and velocity to be $23.80 \pm 0.15$\,d and $1.78 \pm 0.20$\kms, respectively, at a co-latitude of 75.8$^\circ$. We find that the sky-projected angle between the stellar spin axis and the planetary orbital axis is $\lambda = 7.2^{\circ} \pm 3.7^{\circ}$, indicating axial alignment. Our results are consistent with and more precise than published spectroscopic measurements of the Rossiter-McLaughlin effect. These results suggest that WASP-6\,b formed at a much greater distance from its host star and suffered orbital decay through tidal interactions with the protoplanetary disc.
\end{abstract}
\vspace*{-0.8cm}


\section{Introduction}
\label{sec:intro}

At present\footnote{\begin{tt}(http://exoplanet.eu)\end{tt} accessed on 2015/02/20} a total of 1890 planets outside of our own solar system are listed in the authoritative catalogue of \citet{ExoWebPaper}. Of these approximately two thirds have been discovered from  ground-based (e.g.\ SuperWasp: \citealt{SuperWasp}; HAT: \citealt{HAT}) or space-based (CoRoT: \citealt{Corot}; \textit{Kepler}: \citealt{Kepler}) transit surveys, and later confirmed by use of the radial velocity technique \citep{Butler1996,Butler1999,Queloz2000}. Many more candidate exoplanets have been listed in the literature, mainly from the Kepler satellite survey which has also detected several Earth-size planets in the habitable zone (HZ) of their parent star, indicating new worlds with mass and size similar to our own Earth \citep{Borucki2012,Borucki2013}.

During a planetary transit, the planet follows a path (called the transit chord) across the surface of the stellar disc and can be used to probe changes in brightness on the stellar surface \citep{Silva2003}. Starspots have different temperatures to the surrounding photosphere, so emit a different amount of flux. Because photometry measures the change in intensity as a function of time, the occultation of a starspot by the planet causes an anomaly in the light curve \citep{Silva2003}. The anomaly is either an increase or decrease in the amount of light received from the star. If the starspot is a cool spot then the amount of light will increase when the planet crosses the starspot \citep{Rabus2009,Pont2007,Winn2010b}. If the starspot is a hot spot (e.g.\ a facula) then the amount of light will reduce when the planet occults the spot.

At present, when a light curve of a transiting exoplanet is observed to have a starspot anomaly, the transit and the spot are generally modelled separately \citep[e.g][]{Desert2011,Maciejewski2011b,Nutzmann2011,Sanchis2011a}. First, a transit model is fitted to the datapoints not affected by the starspot anomaly. Then the spot-affected residuals versus the best-fitting model are modelled using a Gaussian function \citep[e.g.][]{Sanchis2011a,Sanchis2011b}. This method neglects the fact that the starspot affects the entire transit shape and not just the section where the planet crosses the spot \citep{Ballerini2012}. \citet{Carter2011} use the idea that a starspot on the stellar disc will affect the transit depth to explain the observed changes in transit depth for GJ\,1214. This is due to the change in the star's brightness in its long-term light curve due to starspots rotating on and off the stellar disc.

The transit depth is not the only property of a transit light curve that the starspot affects: it also affects the determination of the measured stellar mean density, stellar radius, orbital inclination and limb darkening (LD) coefficients \citep{Ballerini2012}. The LD coefficients depend on wavelength: because a starspot has a different temperature compared to the surrounding photosphere, it has a different spectral energy distribution and thus different LD coefficients. Therefore the application of a LD law with a single set of coefficients to the entire stellar surface causes a bias in the modelling process \citep{Ballerini2012}. The difference in LD coefficients between the spot and the photosphere can be as much as 30\% in the UV. The effects on the measured stellar radius and orbital inclination of the system are artifacts from errors in the measured planetary radius, which is derived from the transit depth. A change in the measured planetary radius must be compensated for by a change in the measured stellar radius or semimajor axis in order to retain the same transit duration. Starspots can also affect the measured transit midpoint \citep{Sanchis2011a,Barros2013} and create false positives in transit timing measurements. \citet{Sanchis2011a} calculated that a starspot anomaly in a transit of WASP-4 with an amplitude of 0.3 to 0.5\,mmag could produce a timing noise of five to ten seconds.

\subsection{Introducing WASP-6}

The transiting planetary system WASP-6 was discovered by \citet{Gillon2009b} using photometry from the WASP-South telescope. They determined an orbital period of $P = 3.361$\,days for the planet WASP-6\,b. Dedicated photometric observations were then performed in the $i^\prime$ band using the 2-m Faulkes Telescope South (FTS) and in a broad $V$+$R$ band using the RISE instrument \citep{Steele2008} on the 2-m Liverpool Telescope (LT).

Radial velocity (RV) measurements were obtained using two spectrographs: CORALIE on the 1.2-m Euler telescope \citep{Baranne1996,Queloz2000} and HARPS on the ESO 3.6-m telescope \citep{Mayor2003}. \citet{Gillon2009b} determined the stellar mass and radius to be $M_\star = 0.88^{+0.05}_{-0.08}$\Msun\ and $R_\star = 0.870^{+0.025}_{-0.036}$\Rsun, respectively. They found the planetary mass and radius to be $M_{\rm p} = 0.503^{+0.019}_{-0.038}$\Mjup\ and $R_{\rm p} = 1.224^{+0.051}_{-0.052}$\Rjup. They also determined a value for the projected stellar rotational velocity of $v\sin I = 1.4 \pm 1.0$\kms\ from measurements of line widths in the HARPS spectra with a macroturbulence ($v_{\rm mac}$) value of $2$\kms. They noted that if a value of $v_{\rm mac} = 0$\kms\ is used then $v\sin I = 3.0 \pm 0.5$\kms, while if $v_{\rm mac}$ became slightly larger than $2$\kms\ then $v\sin I$ would drop to zero. From their RVs \citet{Gillon2009b} measured the Rossiter-McLaughlin (RM) effect. They found that the system is in alignment with a sky-projected spin orbit alignment, $\lambda = 11^{+14}_{-18}$\,deg.

The spectral analysis of 11 WASP host stars by \citet{Doyle2013} included WASP-6\,A. \citet{Doyle2013} derived new values for the stellar mass and radius of $M_\star = 0.87\pm0.06$\Msun\ and $R_\star = 0.77\pm 0.07$\Rsun, in agreement with those of \citet{Gillon2009b}. \citet{Doyle2013} determined $v_{\rm mac} = 1.4\pm0.3$\kms\ and $v\sin I = 2.4 \pm 0.5$\kms, and an effective temperature of $\Teff = 5375\pm65$\,K.

An optical transmission spectrum for WASP-6 has been constructed using multi-object differential spectrophotometry with the IMACS spectrograph on the Magellan Baade telescope \citep{Jord2013}. The observations comprised 91 spectra covering 480--860\,nm. The analysis yielded a mostly featureless transmission spectrum with evidence of atmospheric hazes and condensates. Most recently, \citet{Nikolov2014} used the {\it Hubble Space Telescope} to perform transmission spectroscopy of WASP-6, and found a haze in the atmosphere of WASP-6\,b. They also determined a rotational modulation of $P_{\rm rot} = 23.6\pm0.5$\,d for WASP-6\,A.


\section{Updates to PRISM \& GEMC}
\label{sec:prism-gemc}

A code written in \textsc{idl}\footnote{The acronym {\sc idl} stands for Interactive Data Language and is a trademark of Exelis Visual Information Solutions. For further details see {\tt http://www.exelisvis.com/ProductsServices/IDL.aspx}.} called \prism\ (Planetary Retrospective Integrated Star-spot Model) was developed to model a starspot anomaly in transit light curves of WASP-19 \citep[see][]{Jeremy2012}. \prism\ uses a pixellation approach to represent the star and planet on a two-dimensional array in Cartesian coordinates. This makes it possible to model the transit, LD and starspots on the stellar disc simultaneously. LD was implemented using the standard quadratic law. \textsc{prism} uses the ten parameters given in Table\,\ref{tab:simresults} to model the system, where the fractional stellar and planetary radii are defined as the absolute radii scaled by the semimajor axis ($r_{\rm \star,p} = R_{\rm \star,p}/a$).

A new optimisation algorithm called \gemc\ (Genetic Evolution Markov Chain) was also created alongside \prism\ to help improve the efficiency of finding a global solution in a rugged parameter space compared to conventional MCMC algorithms \citep{Jeremy2012}. \gemc\ is a hybrid between an MCMC and a genetic algorithm\footnote{A genetic algorithm mimics biological processes by spawning successive generations of solutions based on breeding and mutation operators from the previous generation.} and is based on the Differential Evolution Markov Chain (DE-MC) put forward by \citet{Cajo2006}. During the `burn-in' stage \gemc\ runs $N$ chains in parallel and for every generation each chain is perturbed in a vector towards the current best-fitting chain. Once the burn-in stage has been completed \gemc\ switches to a conventional MCMC algorithm (each chain used in the burn-in begins independent MCMC runs) to determine the parameter uncertainties \citep[see][]{Jeremy2012}.

Since the development of both \prism\ and \gemc\ other authors have used the codes to not only help ascertain the photometric parameters of a transiting system but to also derive the parameters of the starspots observed in transit light curves \citep[e.g.][]{Mancini2013b,Mancini2014,Mohler2013}. \citet{Beky2014} used \prism\ to help calibrate their semi-analytic transit-starspot model \textsc{spotrod}.

Before using both \prism\ and \gemc\ in modelling the WASP-6 system, it was decided to make a few improvements\footnote{The new versions of both \prism\ and \gemc\ are available from http://www.astro.keele.ac.uk/$\sim$jtr}. The original version of \prism\ assumed a circular orbit, as most transiting planets either have a circular orbit or lack a measurement of the orbital eccentricity. However, \citet{Gillon2009b} found that the orbit of WASP-6\,b has a small orbital eccentricity of $e=0.054^{+0.018}_{-0.015}$, with an argument of periastron $\omega = 97.4^{+6.9}_{-13.2}$\,degrees. As a consequence \prism\ was extended to allow for eccentric orbits. $e$ and $\omega$ have been set to roam within the physically bounded ranges of $0 \leq e \leq 1$ and $0^\circ \leq \omega \leq 360^\circ$. A Gaussian prior is used to help constrain the parameter values close to the expected values found in the literature. The logic behind using a Gaussian prior stems from the fact that it is not possible to ascertain these values from photometry alone (due to only observing a small fraction of the orbit) unless an occultation is observed \citep{Kipping2012}. Because we have the knowledge of where the values of $e$ and $\omega$ should lie and that they have an effect on the other system parameters (in particular $i$ and $r_\star$), it is imperative to examine every potential solution selected from a Gaussian probability distribution of $e$ and $\omega$ to accurately estimate the uncertainties in all of the other system parameters.

It was shown by both \citet{Silva2010b} and \citet{Mohler2013} that in some cases there can be more than one starspot anomaly in a single transit light curve. While \prism\ was originally designed to model multiple starspots, the static coding of \gemc\ made it only possible to fit for either a single starspot or a spot free stellar surface. To facilitate further work \gemc\ was modified to fit for multiple starspots. This was accomplished by allowing the initial reading of the input file to be dynamic, so \gemc\ can determine the number of starspots to be fitted based on the number of parameters used. This can be done by adding multiple spot parameter ranges in the input file. It is possible to fix the position of a starspot and therefore assign starspots to sections of the stellar disc where they will not be occulted by the planet, thus allowing investigations of the effects of unocculted starspot on transit light curves.

\prism\ was designed to use the pixellation approach, and to maintain numerical resolution was hard-coded to set the planetary radius at 50 pixels. The host star's radius in pixels was scaled accordingly based on the input parameters. The new version now allows users to set the size of the planetary radius in pixels. This makes it possible to reduce the amount of time required to complete each model iteration, at the cost of numerical resolution (see Section\,\ref{sec:prism_test} for more details). In tests using a planet radius set at 15 pixels it took \prism\ and \gemc\ approximately 13\,s to model a single generation of 256 solutions using synthetic data, which equates to approximately 0.05 seconds per iteration. For comparison, a planetary radius of 50 pixels results in approximately 0.47\,s per iteration.\footnote{These tests were performed on a 2.4\,GHz quad-core laptop.}

To increase the efficiency of determining the parameter uncertainties, the MCMC component of \gemc\ was replaced with DE-MC \citep{Cajo2006}. DE-MC combines the genetic algorithm differential evolution (DE) \citep{Price1997,Storn1997} with MCMC. The combination of DE and MCMC is used to solve a problem in MCMC by determining the orientation and the scale of the step sizes. Adaptive directional sampling in MCMC does solve the orientation problem, but not the scale \citep{Cajo2006}. DE-MC works by creating a population of MCMC chains whose starting points are initialised from overdispersed states and instead of letting the chains run independently and checking for convergence \citep[e.g.][]{Gelman1992} they are instead run in parallel and learn from each other. The perturbation steps taken by each chain are given by Eq.\,\ref{eq:2.8.3a}. Assuming a $d$-dimensional parameter space and using $N$ chains then the population $\textbf{X}$ is a $N \times d$ matrix, with the chains labelled $\textbf{x}_1, \textbf{x}_2, \ldots\ \textbf{x}_N$. Therefore the proposal vector $\textbf{x}_{\rm p}$ is generated by:

\begin{equation}\label{eq:2.8.3a}
\textbf{x}_{\rm p} = \textbf{x}_i + \gamma\left(\textbf{x}_{R1} - \textbf{x}_{R2}\right) + \textbf{e}
\end{equation}

\noindent where $\textbf{x}_i$ is the current $i^{\rm th}$ chain, $\gamma$ is the scale factor calculated from $\gamma = 2.4/\sqrt{2d}$ \citep{Cajo2006}, $\textbf{x}_{R1}$ and $\textbf{x}_{R2}$ are two randomly selected chains and $\textbf{e}$ is drawn from a symmetric distribution with a small variance compared to that of the target. $\textbf{x}_{\rm p}$ is then tested for fitness and if accepted it is used as the next step in $\textbf{x}_i$.

\begin{table*} \centering
\setlength{\tabcolsep}{4pt}
\caption{\label{tab:simresults} Original and recovered parameters from a simulated transit light curve using either
15 or 50 pixels for the planetary radius, plus the interval within which the best fit was searched for using \gemc.}
\begin{tabular}{lcccccc} \hline
Parameter & Symbol            & Original value & Search interval   & Recovered value         & Recovered value         \\
          &                   &                &                   & $r_{\rm p} = 50$\,pixels      & $r_{\rm p} = 15$\,pixels      \\
\hline
Radius ratio                  & $r_{\rm p}/r_\star$ & 0.15  & 0.05 to 0.30      &        0.1496 $\pm$ 0.0013   &        0.1498 $\pm$ 0.0011   \\
Sum of fractional radii       & $r_s + r_{\rm p}$   & 0.25  & 0.10 to 0.50      &        0.2486 $\pm$ 0.0024   &        0.2512 $\pm$ 0.0026   \\
Linear LD coefficient         & $u_1$               & 0.3   & 0.0 to 1.0        &         0.291 $\pm$ 0.104    &        0.281  $\pm$ 0.114    \\
Quadratic LD coefficient      & $u_2$               & 0.2   & 0.0 to 1.0        &         0.192 $\pm$ 0.042    &        0.189  $\pm$ 0.039    \\
Orbital Inclination (degrees)         & $i$                 & 85.0  & 70.0 to 90.0      &         85.16 $\pm$ 0.46     &        85.29  $\pm$ 0.44     \\
Transit epoch (Phase)         & $T_0$               & 0.015 & -0.50 to 0.50     &       0.01494 $\pm$ 0.00011  &       0.01502 $\pm$ 0.00010  \\
Longitude of spot (degrees)   & $\theta$            & 30.0  & -90.0 to +90.0    &         30.50 $\pm$ 1.17     &         30.47 $\pm$ 1.21     \\
Co-latitude of spot (degrees) & $\phi$              & 65.0  & 0.0 to 90.0       &         64.51 $\pm$ 5.83     &         64.17 $\pm$ 5.55     \\
Spot angular radius (degrees) & $r_{\rm spot}$      & 12.0  & 0.0 to 30.0       &         12.73 $\pm$ 2.00     &         12.33 $\pm$ 1.87     \\
Spot contrast                 & $\rho_{\rm spot}$   & 0.8   & 0.0 to 1.0        &         0.797 $\pm$ 0.057    &         0.781 $\pm$ 0.061    \\
\hline \end{tabular} \end{table*}

After the `burn in' stage of a MCMC chain, determining the required step size to allow a 20--25\% acceptance rate can be difficult. For a transit light curve altering the orbital inclination, $i$, by 0.05\% should only cause a small increase in $\chi^2$ but a 0.05\% alteration in the transit midpoint, $T_0$, could cause a large increase in $\chi^2$. DE-MC overcomes the problem with the scale of the step sizes by using the clustering of the chains around the global solution after the `burn in': the difference vector between two randomly selected chains will contain the individual scale for each parameter (e.g.\ 0.05\% for $i$ and 0.00001\% for $T_0$). \citet{Cajo2006} argues that DE-MC is a single $N$-chain that is simply a single random walk Markov Chain in a $N \times d$ dimensional space.

The use of DE-MC in the exoplanet community is increasing, especially for models involving a large number of parameters. For example, models of transiting circumbinary planets can contain over 30 parameters \citep[e.g.][]{Doyle2011,Orosz2012,Welsh2012,Schwamb2013}. To accurately estimate the parameter uncertainties the MCMC component of \gemc\ required $10^6$ function iterations (10 chains each of $10^5$ steps). The DE-MC component requires approximately $2\times10^5$ function iterations (128 chains each of 1500 steps \citep[e.g.][]{Welsh2012}). This equates to a five-fold reduction in the amount of computing time required to fit a transit light curve. When using a set of synthetic transit data it took \gemc\ approximately 5.4\,days to fit the data using a planet radius of 50\,pixels coupled with the MCMC component. The use of a planet radius of 15\,pixels combined with the DE-MC algorithm resulted in \gemc\ taking only 2.7\,hours to fit the same data.


\subsection{Forward simulation of synthetic data}
\label{sec:prism_test}

The modifications to \prism\ and \gemc\ were validated by modelling simulated transit data containing a starspot anomaly. For this test \prism\ was used to create multiple simulated transits with a range of parameters. Noise was then added to the light curves so that the rms scatter between the original simulated light curves and the light curves with added noise was $\approx500$\,ppm. Other levels of noise where also used in similar tests. This was to approximate a realistic level of noise found in transit light curves observed using the defocused photometry technique. Error bars were then assigned to each datapoint to give the original noise-free model a reduced chi squared value of $\chi^2_\nu = 1$.

\begin{figure} \centering \includegraphics[width=0.48\textwidth,angle=0]{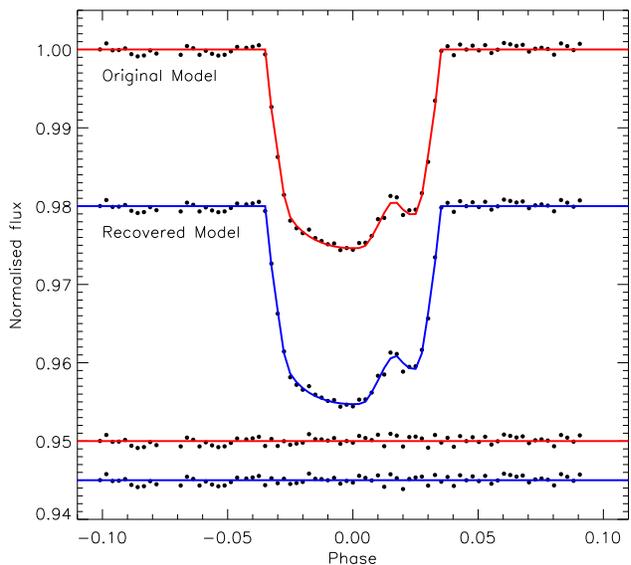}
\caption{\label{fig:simresults} Recovered and original models to simulated transit data created by \prism\ and recovered by \gemc\
and \prism. The residuals are shown at the bottom. The model was calculated with $r_{\rm p} = 50$\,pixels.} \end{figure}

Once a simulated transit light curve had been created, \gemc\ and \prism\ were used in an attempt to recover the initial input parameters. Different values for the planetary pixel radius were also used to test for numerical resolution. Table\,\ref{tab:simresults} shows the results for one of the tests using both $r_{\rm p} = 50$ and $r_{\rm p} = 15$\,pixels, while Fig.\,\ref{fig:simresults} shows the simulated transit light curve together with the original and recovered models for the same test using $r_{\rm p} = 50$\,pixels.

From studying both Table\,\ref{tab:simresults} and Fig.\,\ref{fig:simresults}, it can be seen that the recovered parameter values agree with the original values within their 1$\sigma$ uncertainties. Interestingly, the rms scatter of the recovered model was found to be 499\,ppm while the rms scatter of the original model is 511\,ppm. This showed that \gemc\ not only explored the large parameter search space but also scanned the local area around the global solution to find the best possible fit\footnote{This best fit is in fact a phantom solution generated by the addition of noise.} to the simulated data. This result is expected, and a testament to an optimisation algorithm designed to find the lowest achievable $\chi^2_\nu$ (the recovered solution in this case had a $\chi^2_\nu = 0.94$) in a given parameter space. Similar results were found on all the simulation tests and show that both \gemc\ and \prism\ are capable of accurately and precisely determining the properties of transit light curves.

The recovered parameter values from setting $r_{\rm p} = 50$ and $r_{\rm p} = 15$ pixels also agree within their 1$\sigma$ uncertainties (see Table\,\ref{tab:simresults}). The scale of the 1$\sigma$ uncertainties for when $r_{\rm p} = 15$ are comparable in scale to that of the 1$\sigma$ uncertainties for when $r_{\rm p} = 50$. This indicates that using a smaller number of pixels for the planetary radius (this reduction depends on the number of datapoints and the overall scale of the system being modelled) has little effect on the numerical resolution of the determined parameters or their associated uncertainties. However, using a smaller number of pixels for the planetary radius does affect the smoothness of the plotted best-fit model. It is therefore advisable that, once the best-fitting parameters have been found, \gemc\ is used again with the parameters fixed at the best-fitting values and with $r_{\rm p} = 50$ to calculate a smooth best-fitting model. These tests showed that it is possible to obtain precise results and correctly estimated parameter uncertainties, whilst, using a planetary pixel radius of less than 50. There are, though, some values which should not be used. For example in tests using $r_{\rm p} = 5$ the parameter uncertainties were heavily underestimated, due to numerical noise in the model. By making the planet only 10 pixels across, the numerical resolution decreases to the point where adverse effects can be seen in the results and uncertainties.


\section{Observations and data reduction}
\label{Sec:w6data}

\begin{table*} \centering
\caption{\label{tab:obslogw6} Log of the observations presented for WASP-6. $N_{\rm obs}$ is the number of observations. `Moon illum.' and
'Moon dist.' are the fractional illumination of the Moon, and its angular distance from WASP-6 in degrees, at the midpoint of the transit.}
\begin{tabular}{lcccccccccc} \hline
Date & Start time & End time &$N_{\rm obs}$& Exposure & Filter & Airmass & Moon & Moon & Aperture   & Scatter \\
     &    (UT)    &   (UT)   &             & time (s) &        &         &illum.& dist.& sizes (px) & (mmag)  \\
\hline
2009/06/26 & 06:33 & 10:43 &  91  & 120     & Bessell $R$ & 1.32 $\to$ 1.05 & 0.271 & 160.5 & 65, 90, 110 & 1.215 \\
2009/08/02 & 04:18 & 10:31 &  175 & 90--120 & Bessell $R$ & 1.28 $\to$ 1.44 & 0.934 & 59.6  & 27, 40, 70  & 0.939 \\
2009/08/29 & 02:32 & 07:47 &  129 & 120     & Bessell $R$ & 1.28 $\to$ 1.20 & 0.750 & 63.8  & 28, 40, 60  & 0.598 \\
2010/07/31 & 03:51 & 10:20 &  193 & 80      & Bessell $R$ & 1.45 $\to$ 1.34 & 0.686 & 42.4  & 25, 35, 55  & 0.591 \\
\hline \end{tabular} \end{table*}

Four transits of WASP-6 were observed on 2009/06/26, 2009/08/02, 2009/08/29 and 2010/07/31 by the MiNDSTEp consortium \citep{Dominik2010} using the Danish 1.54\,m telescope at ESO's La Silla observatory in Chile. The instrument used was the DFOSC imager, operated with a Bessell $R$ filter. In this setup the CCD covers a field of view of $(13.7\am)^{2}$ with a pixel scale of 0.39\as\,pixel$^{-1}$. The images were unbinned but windowed for faster readout, resulting in a dead time between consecutive images of between 22 and 35\,s. The exposure times were 80--120\,s. The Moon's brightness and distance to the target star is given in Table\,\ref{tab:obslogw6}. The telescope was defocused and autoguiding was maintained through all observations. The amount of defocus applied caused the resulting PSFs to have a diameter of 86 pixels for the night of 2009/06/26, 32 pixels for the night of 2009/08/02, 44 pixels for the night of 2009/08/29 and 37 pixels for the night of 2010/07/31.

\begin{figure} \centering \includegraphics[width=0.48\textwidth,angle=0]{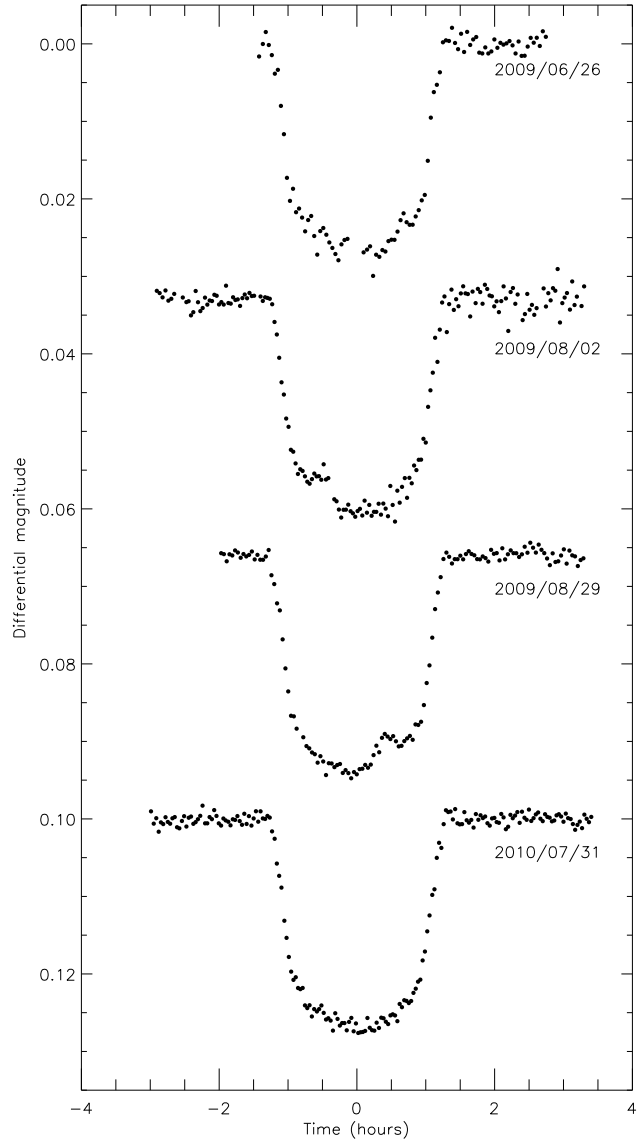}
\caption{\label{fig:data} The four light curves of WASP-6 presented in this work, in the order presented
in Table\,\ref{tab:obslogw6}. Times are given relative to the midpoint of each transit.} \end{figure}

We reduced the data in an identical fashion to \citet{Sou2009,Sou2009b}. In short, aperture photometry was performed with an {\sc idl} implementation of {\sc daophot} \citep{Stetson1987}, and the aperture sizes were adjusted to obtain the best results (see Table\,\ref{tab:obslogw6}). A first order polynomial was then fitted to the outside-transit data whilst simultaneously optimising the weights of the comparison stars. The resulting data have scatters ranging from 0.591 to 1.215 mmag per point versus a transit fit using \textsc{prism}. The timestamps from the fits files were converted to BJD/TDB. An observing log is given in Table\,\ref{tab:obslogw6} and the final light curves are plotted in Fig.\ref{fig:data}.


\section{Data analysis}
\label{Sec:w6results}

\begin{table*}
\setlength{\tabcolsep}{4pt}
\caption{\label{tab:resultsw6} \small Derived photometric parameters from each light curve, plus the interval within which the best fit was searched for using \gemc.}
\begin{tabular}{lccccc} \hline
Parameter                     & Search interval   & 2009/06/26                  & 2009/08/02                  & 2009/08/29                  & 2010/07/31                  \\
\hline
Radius ratio                  & 0.05 to 0.30      &        0.1443 $\pm$ 0.0055  &        0.1444 $\pm$ 0.0043  &        0.1474 $\pm$ 0.0017  &        0.1454 $\pm$ 0.0021  \\
Sum of fractional radii       & 0.10 to 0.50      &        0.1102 $\pm$ 0.0060  &        0.1109 $\pm$ 0.0048  &        0.1115 $\pm$ 0.0025  &        0.1114 $\pm$ 0.0023  \\
Linear LD coefficient         & 0.0 to 1.0        &         0.366 $\pm$ 0.119   &         0.397 $\pm$ 0.116   &         0.368 $\pm$ 0.077   &         0.402 $\pm$ 0.067   \\
Quadratic LD coefficient      & 0.0 to 1.0        &         0.245 $\pm$ 0.191   &         0.325 $\pm$ 0.222   &         0.186 $\pm$ 0.123   &         0.192 $\pm$ 0.134   \\
Orbital Inclination (degrees)         & 70.0 to 90.0      &         88.47 $\pm$ 0.99    &         88.55 $\pm$ 0.85    &         88.33 $\pm$ 0.48    &         88.36 $\pm$ 0.53    \\
Transit epoch (BJD/TDB)       & $\pm$0.5 in phase & 2455009.83622 $\pm$ 0.00021 & 2455046.80720 $\pm$ 0.00015 & 2455073.69529 $\pm$ 0.00013 & 2455409.79541 $\pm$ 0.00010 \\
Longitude of spot (degrees)   & -90 to +90        &                             &        -26.15 $\pm$ 1.52    &         21.30 $\pm$ 0.99    &                             \\
Co-latitude of Spot (degrees) & 0.0 to 90.0       &                             &         78.76 $\pm$ 1.58    &         72.77 $\pm$ 1.12    &                             \\
Spot angular radius (degrees) & 0.0 to 30.0       &                             &         12.25 $\pm$ 1.40    &         12.17 $\pm$ 0.81    &                             \\
Spot contrast                 & 0.0 to 1.0        &                             &         0.649 $\pm$ 0.187   &         0.798 $\pm$ 0.082   &                             \\
\hline \end{tabular}
\end{table*}

All four transits were modelled using \prism\ and \gemc. To do this a large parameter search space was selected to allow the global best fit solution to be found. As discussed in \citet{Jeremy2012}, the ability of \gemc\ to find the global minimum in a short amount of computing time meant that it was possible to search a large area of parameter space to avoid the possibility of missing the best solution. The parameter search ranges used in analysing the WASP-6 datasets are given in Table\,\ref{tab:resultsw6}. We modelled the two datasets containing a starspot anomaly independently, in order to obtain two sets of starspot parameters. This helps the investigation of whether the two anomalies are due to the same starspot (see Section\,\ref{Sec:WASP-6introstarspot}).

\begin{figure} \includegraphics[width=0.48\textwidth,angle=0]{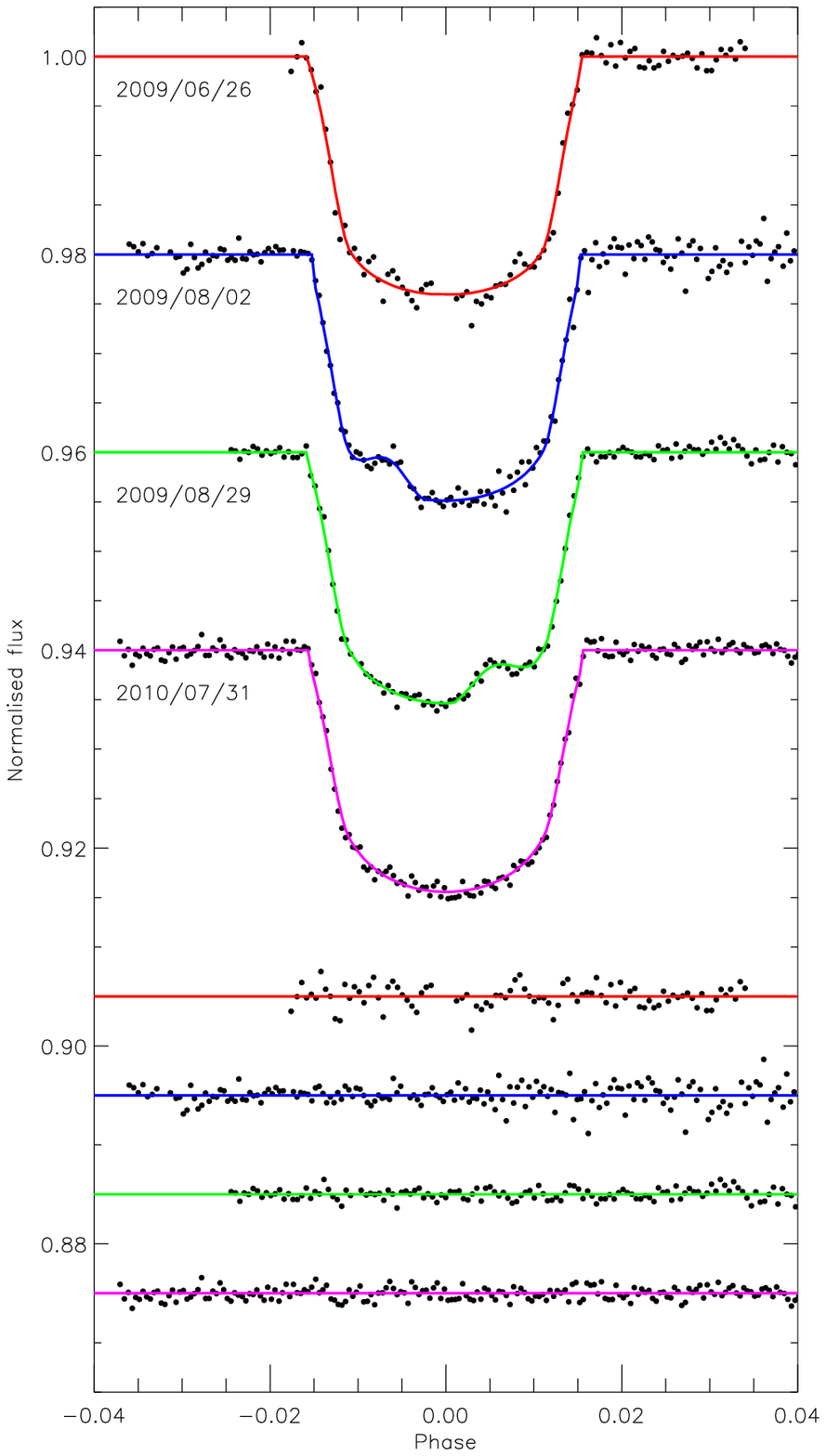}
\caption{\label{fig:w6lc} Transit light curves and the best-fitting models of
WASP-6. The residuals are displayed at the base of the figure.} \end{figure}

The separate models of the four datasets of WASP-6 have parameters which are within 1$\sigma$ of each other (Table\,\ref{tab:resultsw6}). \citet{Ballerini2012} noted that starspots can affect the LD coefficients by up to 10\% in the $R$ band. This is not seen in the WASP-6 data, unlike in the transit data of WASP-19 \citep{Jeremy2012}. The scatter around the weighted mean is $\chi^2_\nu = 0.149$ for the linear coefficient and $0.355$ for the quadratic coefficient. The error bars on the LD coefficients are too large to allow the effects of starspots to be detected. This is due to the lower quality of the data compared to WASP-19 \citep{Jeremy2012}. The combined best-fit LD coefficients are also in agreement within their 1$\sigma$ uncertainties with the theoretically predicted values for WASP-6\,A of $u_1 = 0.4125$ and $u_2 = 0.2773$ \citep{Claret2000}.


\subsection{Photometric results}
\label{Sec:w6photoresults}

\begin{table}
\caption{\label{tab:resultsw6final} Combined system and spot parameters for WASP-6.
The system parameters are the weighted means from all four data sets.
The spot angular size and contrast are the weighted means from the two transits containing a starspot anomaly.}
\setlength{\tabcolsep}{4pt}
\begin{tabular}{lcc} \hline
Parameter & Symbol & Value   \\
\hline
Radius ratio                             & $r_{\rm p}/r_\star$         & 0.1463 $\pm$ 0.0012 \\
Sum of fractional radii                  & $r_s + r_{\rm p}$       & 0.1113 $\pm$ 0.0015 \\
Linear LD coefficient                    & $u_1$             &  0.386 $\pm$ 0.043  \\
Quadratic LD coefficient                 & $u_2$             &  0.214 $\pm$ 0.077  \\
Orbital Inclination (degrees)                    & $i$               &  88.38 $\pm$ 0.31   \\
Spot angular radius (degrees)            & $r_{\rm spot}$    &  12.19 $\pm$ 0.70   \\
Spot contrast                            & $\rho_{\rm spot}$ &  0.774 $\pm$ 0.075  \\
Stellar rotation period (d)              & $P_{\rm rot}$     &  23.80 $\pm$ 0.15   \\
Projected spin orbit alignment (degrees) & $\lambda$         &    7.2 $\pm$ 3.7    \\
\hline \end{tabular} \end{table}

The final photometric parameters for the WASP-6 system are given in Table\,\ref{tab:resultsw6final} and are weighted means together with their 1$\sigma$ uncertainties of the results from the four individual fits. Fig.\,\ref{fig:w6lc} compares the light curves to the best-fitting models, including the residuals.

\begin{table} \begin{center}
\caption{\label{tab:minimaw6} Times of minimum light of WASP-6
and their residuals versus the ephemeris derived in this work.
\newline {\bf References:}
(1) \citet{Gillon2009b};
(2) This work;
(3) \citet{Dragomir2011};
(4) \citet{Jord2013};
(5) \citet{Sada2012};
(6) \citet{Nikolov2014}}
\vspace{0.5cm}
\begin{tabular}{l@{\,$\pm$\,}l r r c} \hline
\multicolumn{2}{l}{Time of minimum}   & Cycle  & Residual & Reference \\
\multicolumn{2}{l}{(BJD/TDB $-$ 2400000)} & no.    & (BJD)    &           \\
\hline
54425.02167 & 0.00022 &     0.0 & -0.00013 &  1 \\   
55009.83622 & 0.00021 &   174.0 &  0.00006 &  2 \\   
55046.80720 & 0.00015 &   185.0 &  0.00001 &  2 \\   
55073.69529 & 0.00013 &   193.0 &  0.00008 &  2 \\   
55409.79541 & 0.00010 &   293.0 & -0.00000 &  2 \\   
55446.76621 & 0.00058 &   304.0 & -0.00023 &  3 \\   
55473.65438 & 0.00016 &   312.0 & -0.00007 &  4 \\   
55846.72540 & 0.00045 &   423.0 & -0.00028 &  5 \\   
56088.71800 & 0.00013 &   495.0 &  0.00017 &  6 \\   
56095.43973 & 0.00017 &   497.0 & -0.00011 &  6 \\   
56132.41081 & 0.00010 &   508.0 & -0.00005 &  6 \\   
\hline \end{tabular} \end{center} \end{table}

\begin{figure*} \includegraphics[width=\textwidth,angle=0]{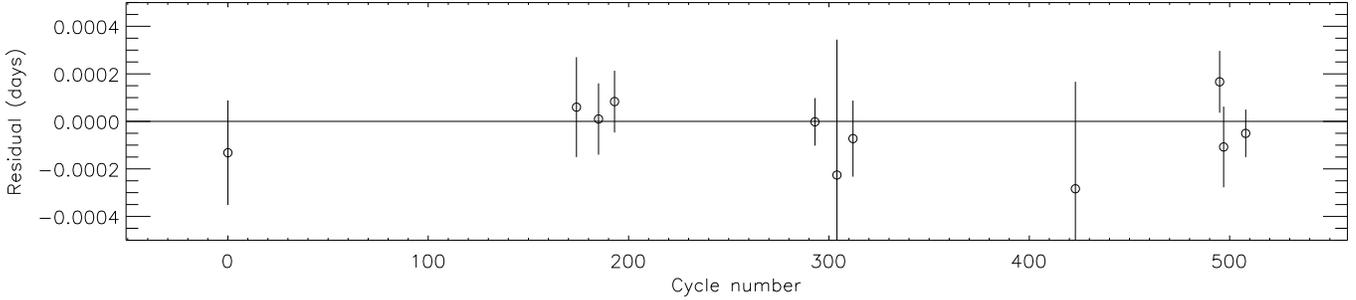}
\caption{\label{fig:ocw6} Residuals of the available times of mid-transit
versus the orbital ephemeris found for WASP-6. The four timings from
this work are the cluster of three points between the cycle numbers
170--200 and the point close to cycle 290.} \end{figure*}

The available times of mid-transit for WASP-6 were collected from the literature \citep{Gillon2009b,Dragomir2011,Sada2012,Jord2013,Nikolov2014}. All timings were converted to the BJD/TDB timescale and used to obtain an improved orbital ephemeris: $$ T_0 = {\rm BJD/TDB} \,\, 2\,454\,425.02180 (11) \, + \,3.36100208 (31) \times E $$ where $E$ represents the cycle count with respect to the reference epoch and the bracketed quantities represent the uncertainty in the final two digits of the preceding number. Fig.\,\ref{fig:ocw6} and Table\,\ref{tab:minimaw6} show the residuals of these times against the ephemeris. The results show no evidence for transit timing variations.

Initially we used the quoted mid-transit time from \citet{Gillon2009b}, but found that this value disagreed with the other 10 mid-transit times at the 2.2$\sigma$ level. This may be because the value found by \citet{Gillon2009b} was derived by simultaneously fitting the original WASP data plus two incomplete transits from RISE and a single complete transit from the FTS. We therefore used the same approach as \citet{Nikolov2014} and fitted (using \prism) the archival FTS light curve to determine the mid-transit time. The value found using just the FTS data is in better agreement (0.6$\sigma$) with the other ten mid-transit times. Therefore it was decided to use the mid-transit time from the FTS light curve in our analysis, not just due to the better agreement but also due to the fact that it comes directly from a light curve covering a full transit.


\subsection{Physical properties of the WASP-6 system}
\label{Sec:w6phyiscalresults}

With the photometric properties of WASP-6 measured the physical characteristics could be determined. The analysis followed the method of \citet{Me09mn}, which uses the parameters measured from the light curves and spectra, plus tabulated predictions of theoretical models. We adopted the values of $i$, $r_{\rm p}/r_\star$ and $r_\star + r_{\rm p}$ from Table\,\ref{tab:resultsw6final}, the orbital velocity amplitude $K_{\star} = 74.3^{+1.7}_{-1.4}$\ms\ and eccentricity $e = 0.054^{+0.018}_{-0.015}$ from \citet{Gillon2009b}, and the stellar effective temperature $\Teff = 5375\pm65$\,K and metal abundance $\FeH = -0.15 \pm 0.09$ from \citet{Doyle2013}.

An initial value of the velocity amplitude of the planet, $K_{\rm p}$, was used to calculate the physical properties of the system using standard formulae and the physical constants listed by \citet{Me11mn}. The mass and \FeH\ of the star were then used to obtain the expected \Teff\ and radius, by interpolation within a set of tabulated predictions from theoretical stellar models. $K_{\rm p}$ was iteratively refined until the best agreement was found between the observed and expected \Teff, and the measured $r_{\rm \star}$ and expected $\frac{R_{\rm \star}}{a}$. This was performed for ages ranging from the zero-age to the terminal-age main sequence, in steps of 0.01\,Gyr. The overall best fit was found, yielding estimates of the system parameters and the evolutionary age of the star.

This procedure was performed separately using five different sets of stellar theoretical models \citep[see][]{Me10mn}, and the spread of values for each output parameter was used to assign a systematic error. Statistical errors were propagated using a perturbation algorithm \citep[see][]{Me10mn}.

The final results of this process are in reasonable agreement with themselves and with published results for WASP-6. The final physical properties are given in Table\,\ref{tab:modelw6} and incorporate separate statistical and systematic errorbars for those parameters which depend on the theoretical models. The final statistical errorbar for each parameter is the largest of the individual ones from the solutions using each of the five different stellar models. The systematic errorbar is the largest difference between the mean and the individual values of the parameter from the five solutions.

\begin{table} \centering
\caption{\label{tab:modelw6} Physical properties of the WASP-6 system. Where two errorbars are
given, the first is the statistical uncertainty and the second is the systematic uncertainty.}
\begin{tabular}{l r@{\,$\pm$\,}c@{\,$\pm$\,}l}
\hline
Parameter & \mcc{Value} \\
\hline
$M_{\rm A}$    (\Msun) & 0.836    & 0.063    & 0.024       \\
$R_{\rm A}$    (\Rsun) & 0.864    & 0.024    & 0.008       \\
$\log g_{\rm A}$ (cgs) & 4.487    & 0.017    & 0.004       \\
$\rho_{\rm A}$ (\psun) & \mcc{$1.296 \pm 0.053$}           \\
$M_{\rm b}$    (\Mjup) & 0.485    & 0.027    & 0.009       \\
$R_{\rm b}$    (\Rjup) & 1.230    & 0.035    & 0.012       \\
$g_{\rm b}$     (\mss) & \mcc{$7.96 \pm 0.30$}             \\
$\rho_{\rm b}$ (\pjup) & 0.244    & 0.014    & 0.002       \\
\Teq\              (K) & \mcc{$1184 \pm   16$}             \\
\safronov\             & 0.0390   & 0.0014   & 0.0004      \\
$a$               (AU) & 0.0414   & 0.0010   & 0.0004      \\
Age              (Gyr) & \ermcc{9.0}{8.0}{12.7}{4.0}{9.0}  \\
\hline \end{tabular}  \end{table}


\section{Starspot anomalies}
\label{Sec:WASP-6introstarspot}

Two of the light curves, from 2009/08/02 and 2009/08/29, contain apparent starspot anomalies (see Fig.\,\ref{fig:data}). Due to a 27 day gap between the two light curves it is not possible to conclusively demonstrate that the anomalies are due to the same spot. But if so, the stellar rotation period and sky-projected spin orbit alignment can be calculated and compared to the values found by \citet{Gillon2009b}, \citet{Doyle2013} and \citet{Nikolov2014}. This will allow an indirect check on whether the two spot anomalies are due to the same starspot.

Firstly we consider whether the spot could last for a 27 day period. On the Sun a spot's lifetime $T$ is proportional to its size $A_0$ following the Gnevyshev-Waldmeier (G-W) Relation \citep{Gnevyshev1938,Waldmeier1955}.

\begin{equation}\label{eq:4.1}
 A_0 = WT
\end{equation}

\noindent where $A_0$ is measured in MSH (micro-Solar hemispheres) and $T$ is in days. \citet{Petrovay1997} state that $W=10.89\pm0.18$\,MSH day$^{-1}$. \citet{Henwood2010} showed that large sunspots also followed the Gnevyshev-Waldmeier relationship. If the same relationship is applied to starspots then a minimum lifetime of 30\,days requires a minimum size of 327\,MSH, or an angular radius of just greater than $1^\circ$. \citet{Bradshaw2014} argues that the standard solar G-W relation overestimates the lifetime of a starspot. \citet{Bradshaw2014} uses turbulent magnetic diffusivity at supergranule size scales to calculate the magnetic diffusivity which in turn allows $W$ in the G-W relation to be re-calculated. Depending on the turbulent scale length being used and to have a minimum lifetime of 30\,days requires a angular radius of $3^\circ$ to $9^\circ$ \citep[see fig.\,1][]{Bradshaw2014}. The sizes of the starspot anomalies in the WASP-6 light curves are greater than $10^\circ$, so we conclude that a single spot can last sufficiently long to cause both anomalies, irrespective of the turbulent scale length used.

\subsection{Starspot anomalies results}
\label{Sec:WASP-6-spots}

\begin{figure}
\includegraphics[width=0.24\textwidth,angle=0]{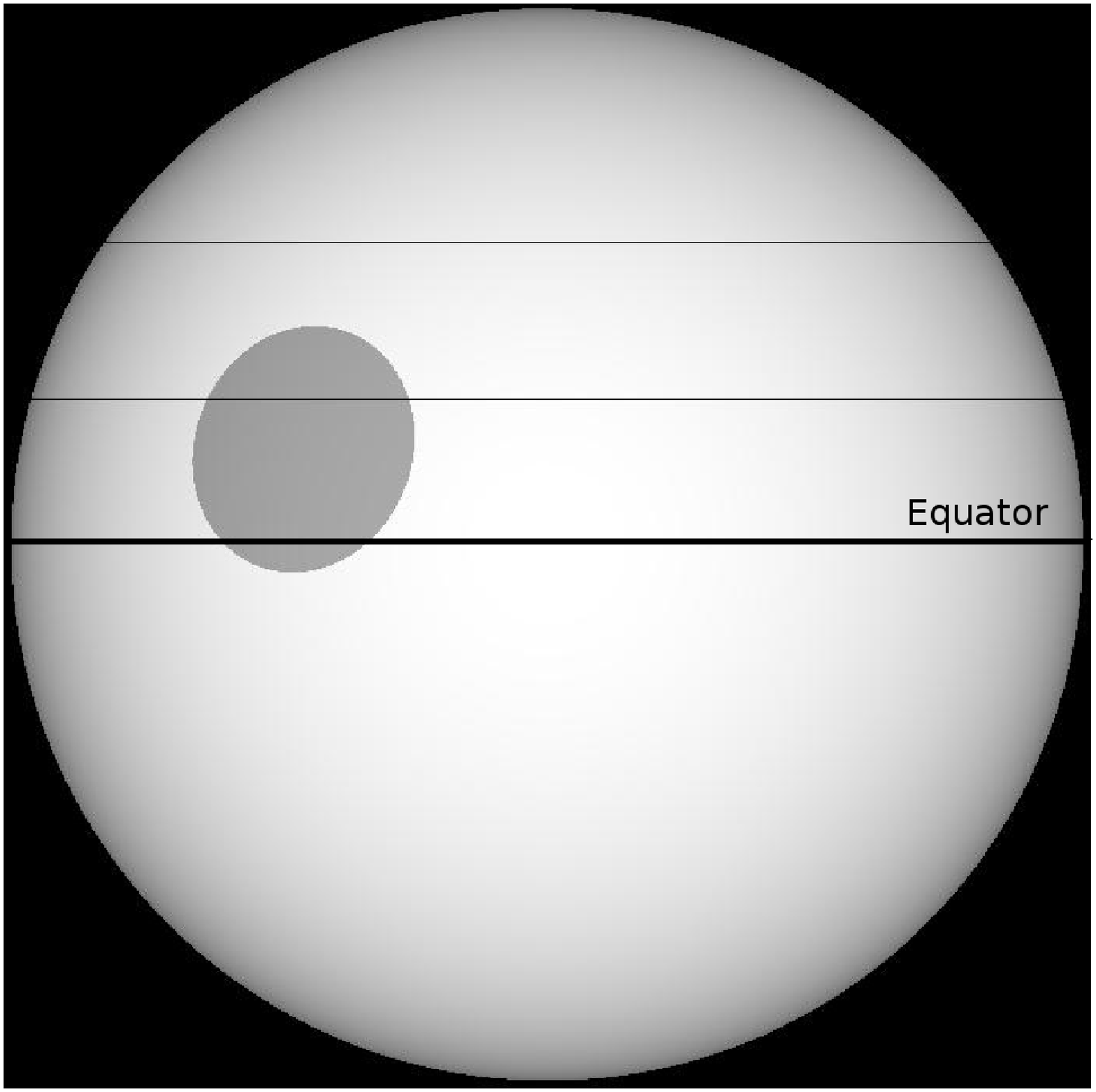}
\includegraphics[width=0.24\textwidth,angle=0]{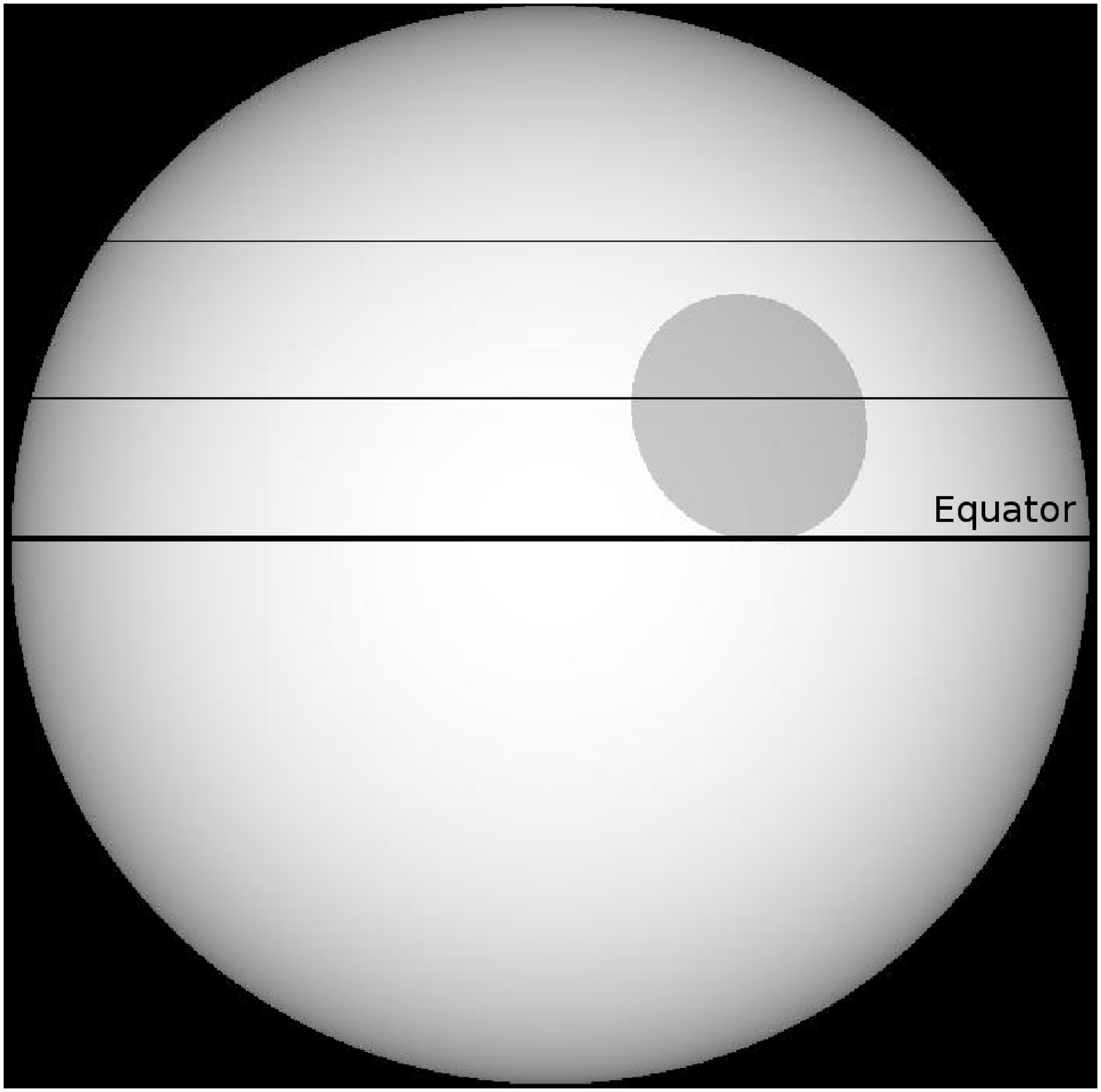}
\caption{\label{fig:spot1w6} Representation of the stellar disc, starspot, transit chord and equator for the two datasets of
WASP-6 containing spot anomalies. The axis of stellar rotation lies in the plane of the page and in the case of $\lambda = 0^\circ$ points upwards.} \end{figure}

The results from modelling the two spot anomalies suggest that they are due to the same spot rotating around the surface of the star, as the spot sizes and contrasts are in good agreement and the lifetime of a spot this size is much greater than the time interval between the two spotted transits. Fig.\,\ref{fig:spot1w6} is a representation of the stellar disc, the spot and the transit chord for the two nights of observations.

By assuming that the two spot anomalies are indeed caused by the same spot, it is straightforward to calculate the sky-projected spin orbit alignment of the system. We find a value of $\lambda = 7.2^{\circ} \pm 3.7^{\circ}$ from the measured positions of the starspot during the two transits.

It is also possible to calculate the rotational period of the star, using the spot positions and an estimate of the number of stellar rotations which occurred between the two transits \citep[see][]{Jeremy2012,Mancini2014}. Due to the 27 day gap between the light curves the star could have rotated $N$ full rotations plus $47.5^{\circ} \pm 2.5^{\circ}$. If $N=0$ then this would imply that WASP-6 has a rotation period of approximately 200\,days, which is extremely long for a main sequence G-star. If $N=1$ then the spot has travelled $407.5^{\circ} \pm 2.5^{\circ}$ between the transits, giving a rotational period of $P_{\rm rot} = 23.80\pm0.15$\,d at a co-latitude of $75.8^{\circ}$. This is in excellent agreement with the measurement of $P_{\rm rot} = 23.6\pm0.5$\,d from \citet{Nikolov2014}. Combining this with the stellar radius (see Table\,\ref{tab:modelw6}), the latitudinal rotational velocity of the star was calculated to be $v_{\left(75.8^\circ\right)} = 1.78\pm0.20$\kms. This is also in agreement with $v \sin I$ from both \citet{Gillon2009b} and \citet{Doyle2013}. If $N=2$ then the spot has travelled $767.5^{\circ} \pm 2.5^{\circ}$, giving a rotational period of $P_{\rm rot} = 12.63\pm0.15$\,d at a co-latitude of $75.8^{\circ}$ (or $v_{\left(75.8^\circ\right)} = 3.36\pm0.20$\kms). This agrees with the $v \sin I$ from \citet{Gillon2009b} and \citet{Doyle2013}, but not with the $P_{\rm rot}$ from \citet{Nikolov2014}. The agreement with \citet{Gillon2009b} and \citet{Doyle2013} is due to the fact that any value of $v$ that is found to be greater than $v \sin I$ can be considered to agree based on the nature of $\sin I$. We conclude that the $N=1$ case is much more likely than the two alternatives discussed above.

\subsection{Degeneracy of the stellar rotation period}
\label{Sec:WASP-6-rotation}

Whilst there is no clear photometric signal in the SuperWASP light curve of WASP-6, \citet{Nikolov2014} were able to measure a rotation period of $P_{\rm rot} = 23.6\pm0.5$\,d from photometry of higher precision, however, none of the STIS observations detected a starspot anomaly indicating that starspots on WASP-6\,A are either rare or of low contrast. This is also supported by the upper limit of the photometric variability of about 1\% \citep{Nikolov2014}. There are also two measurements of $v \sin I$ from \citet{Gillon2009b} ($v\sin I = 1.4 \pm 1.0$\kms) and \citet{Doyle2013} ($v\sin I = 2.4 \pm 0.5$\kms). Both $v \sin I$ measurements agree with the $v$ found when combining $P_{\rm rot}$ and $R_\star$ at a co-latitude of $75.8^{\circ}$ to give either $v_{\left(75.8^\circ\right)} = 1.78\pm0.20$\kms\ or $v_{\left(75.8^\circ\right)} = 3.36\pm0.20$\kms. The problem that arises from checking measurements of $v$ against $v \sin I$ is that due to the $\sin I$ projection factor any value for $v$ that is found to be greater than $v \sin I$ can be considered to agree. A second unknown is the amount of differential rotation that is experienced by WASP-6\,A. In the absence of any differential rotation the single full rotation value of $P_{\rm rot} = 23.80\pm0.15$\,d would lead to an equatorial rotational velocity of $v = 1.84\pm0.20$\kms. This result agrees again with the $v \sin I$ value from both \citet{Gillon2009b} and \citet{Doyle2013}. Our results from \prism\ do show though that the two starspot positions are only approximately $10^{\circ}$ from the stellar equator. As such the effect from differential rotation would be small, so any large divergence of $v$ from $v \sin I$ would imply that $I \ll 90^\circ$.

WASP-6\,A has $\Teff = 5375\pm65$\,K \citep{Doyle2013} so is a cool star ($\Teff<6250$\,K). The trend seen between host star \Teff s and projected orbital obliquity (see Fig.\,\ref{fig:con2}) suggests that the orbital rotation axis of WASP-6\,b should be aligned with the stellar rotation axis of WASP-6\,A. For this to be true then $I$ would have to be $\approx 90^\circ$, and thus $\sin I \approx 1$. If this is the case then the value $v_{\left(75.8^\circ\right)} = 3.36\pm0.20$\kms\ no longer agrees with the $v \sin I$ from either \citet{Gillon2009b} or \citet{Doyle2013}. This supports the supposition that the rotation period of WASP-6\,A is $P_{\rm rot} = 23.80\pm0.15$\,d. \citet{Brown2014} calculated the stellar rotation period of WASP-6\,A to be $P_{\rm rot} = 27.1^{+3.6}_{-3.8}$\,d from Gaussian distribution sampling of $v\sin I$, $i$ and $R_A$, which is further evidence for the conclusion that the rotation period of WASP-6\,A is $P_{\rm rot} = 23.80\pm0.15$\,d.


\section{Discussion and conclusions}
\label{Sec:w6conculsions}

We have determined the physical properties of the WASP-6 planetary system (Table\,\ref{tab:modelw6}) based on four new high-precision transit light curves, finding values which are consistent with and more precise than those in the literature. We find the mass and radius of the host star to be $0.836\pm 0.063\Msun$ and $0.864\pm0.024\Rsun$, respectively. For the planet we find a mass of $0.485\pm 0.027\Mjup$, a radius of $1.230\pm0.035\Rjup$ and a density of $0.244\pm0.014\pjup$. These results also serve as a secondary check for the accuracy of the \prism\ and \gemc\ codes. By studying the individual results for each of the four transits (see Table\,\ref{tab:resultsw6}) it can be seen that the system parameters from each light curve agree within their 1$\sigma$ uncertainties. This shows that \prism\ can retrieve reliable photometric properties from transit light curves containing starspot anomalies.

The four transits of WASP-6 were modelled using \prism\ and \gemc. Two of the transits contained a starspot anomaly but are separated by 27\,days. Whilst it is not possible to prove that the two spot anomalies are caused by the same starspot, the available evidence strongly favours this scenario. The results from \prism\ show that the angular size and contrast of the starspot in both light curves agree to within 0.05$\sigma$ and 0.73$\sigma$, respectively. As with WASP-19 \citep[see][]{Jeremy2012}, only part of the starspot(s) is on the transit chord (Fig.\,\ref{fig:spot1w6}). Because the light curve only holds information on what is happening inside the transit chord then a likely scenario is that the planet is passing over a band of smaller starspots which form an active region on WASP-6. In this active region, there could be a number of starspots each with sizes much less than $1^\circ$ and therefore lifetimes shorter than 30\,days (see Section\,\ref{Sec:WASP-6introstarspot}). Future observations may allow changes to be seen in the overall contrast from the starspot region. In either case as a whole the region would remain a similar size and shape over a 27\,day period.

In the case of a single large starspot, $r_{\rm spot}$ (Table\,\ref{tab:resultsw6final}) and $R_\star$ (Table\,\ref{tab:modelw6}) can be combined to find the starspot radius. We find $R_{\rm spot} = 127902\pm11102$\,km, which equates to approximately 4.5\% of the visible stellar surface. This value is similar to starspots found on other G-type stars \citep{Strassmeier2009}.

If the two starspot anomalies are assumed to be generated by the planet crossing the same starspot then it is possible to calculate the latitudinal rotation period of WASP-6. It was found that either $P_{\rm rot} = 23.80\pm0.15$\,d or $P_{\rm rot} = 12.63\pm0.15$\,d  at a co-latitude of $75.8^{\circ}$. These calculations assumed that WASP-6 had made either one or two full rotations prior to the difference seen in the light curves.

Even without knowing the number of full rotations that WASP-6 completed between the two spotted light curves, if the starspot anomalies are due to the same spot then the sky-projected spin orbit alignment $\lambda$ of the system can be measured. We find $\lambda = 7.2^{\circ} \pm 3.7^{\circ}$. This result agrees with, and is more precise than, the previous measurement of $\lambda$ using the RM effect ($\lambda = 11^{\circ}$\,$^{+14}_{-18}$; \citealt{Gillon2009b}). $\lambda$ gives the lower boundary of the true spin-orbit angle, $\psi$. As stated by \citet{Fabrycky2009}, finding a small value for $\lambda$ can be interpreted in different ways. Either $\psi$ lies close to $\lambda$ and the system is aligned, or $\psi$ lies far from $\lambda$ and the system is not aligned. As discussed in Section\,\ref{Sec:WASP-6-spots} because the spot is close to the stellar equator then it could be assumed that the change in $v$ at the equator due to differential rotation would be small. Coupled with the uncertainties measured in $v \sin I$ from both \citet{Gillon2009b} and \citet{Doyle2013} it is plausible that $\sin I \approx 1$ and therefore $\psi \approx 7^\circ$ if $P_{\rm rot} = 23.80\pm0.15$\,d. As a consequence we have two different scenarios: an aligned system with a slowly rotating star or a misaligned system with a rapidly rotating star. Taking into account the \Teff\ of WASP-6\,A and the statistical trend seen in misaligned systems it is more probable that the WASP-6 system is in fact aligned, suggesting $\psi \approx 7^\circ$ and $P_{\rm rot} = 23.80\pm0.15$\,d. It would be desirable to observe consecutive transits of WASP-6 in an attempt to definitively identify multiple planetary crossings of a single starspot and to precisely determine $P_{\rm rot}$, $\lambda$ and potentially $\psi$ of WASP-6.

If the starspot anomalies are due to the same starspot, $\lambda = 7.2^{\circ} \pm 3.7^{\circ}$ and there is no direct evidence for a spin-orbit misalignment in the WASP-6 system. With potentially a low obliquity and a cool host star, WASP-6 seems to follow the idea put forward by \citet{Winn2010c} that planetary systems with cool stars will have a low obliquity. It also lends weight to the idea that WASP-6\,b formed at a much greater distance from its host star and suffered orbital decay through tidal interactions with the protoplanetary disc (i.e.\ either Type\,I or Type\,II disc-migration, \citealt{Ward1997}).

\begin{figure} \includegraphics[width=0.48\textwidth,angle=0]{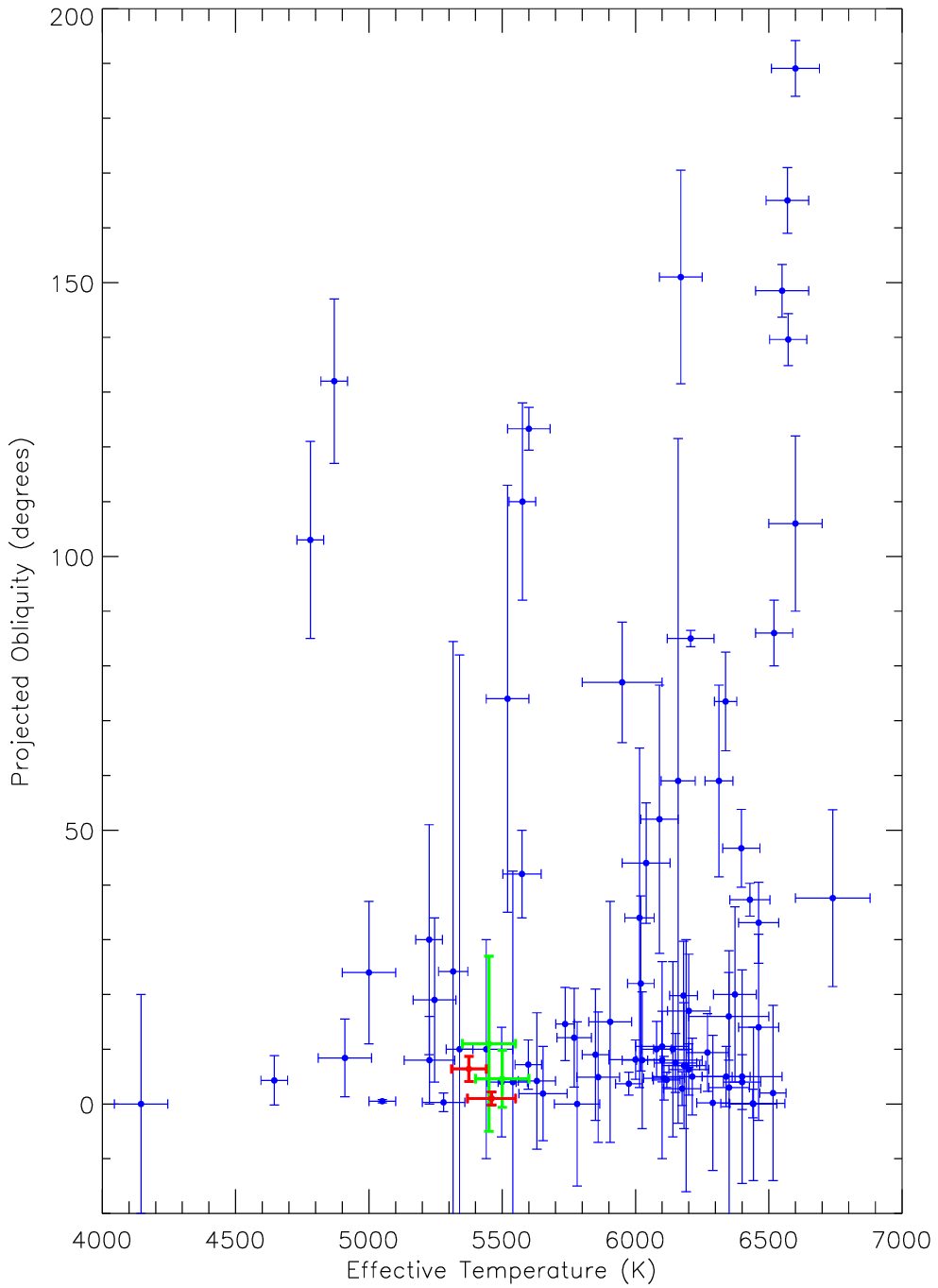}
\caption{\label{fig:con2} $|\lambda|$ against \Teff\ for 83 transiting planets from TEPCat including WASP-19 and WASP-6.
The green and red datapoints are WASP-6 (left) and WASP-19 (right). The green datapoints represent values from the
literature (WASP-6: \citealt{Gillon2009b}; WASP-19: \citealt{Hellier2011}) and the red datapoints represent the values found
from this work and \citet{Jeremy2012}. The trend in the data suggests that cool host stars harbour aligned systems.} \end{figure}

At present there are 83 transiting planets with published $\lambda$ values\footnote{All measured $\lambda$ and \Teff\ values of the known planetary systems were obtained from the 2014/10/20 version of the TEPCat catalogue \citep{Me11mn}. {\tt \newline (http://www.astro.keele.ac.uk/$\sim$jkt/tepcat/)}}. The $\lambda$ values for WASP-6 (this work) and WASP-19 \citep{Jeremy2012} were updated and a plot of $\lambda$ against \Teff\ was created (see Fig.\,\ref{fig:con2}). To remove any ambiguity in the plot due to negative values of $\lambda$, we plot its absolute value. It can be seen that a large proportion (75\,\%) of cool stars ($\Teff<6250$\,K) are in aligned systems, while the majority (56\,\%) of hot host stars have misaligned systems. This trend supports \citet{Winn2010c} in that cool stars with hot Jupiters will have low obliquities. This trend can also be explained by the time required for the system to align. Hot stars will have thinner convective zones and will therefore take longer to align the photosphere with the planetary orbit. Because of this, by examining $\lambda$ of hot stars a greater proportion will have misaligned systems compared to cool stars where the alignment process is much shorter and so will have a higher proportion of aligned systems. Cool stars also live longer so the ones that are observed are on average older. They have therefore had more time for tidal effects to work \citep{Triaud2011}.

By determining $\lambda$ and $\psi$ of the planetary system it is possible to begin to understand the primary process in the dynamical evolution of the system. The RM effect can be used to ascertain a value for $\lambda$. One limitation of this method though is from an excess RV jitter (stellar activity e.g.\ starspots). Therefore, the use of the RM effect either requires magnetically quiet stars or the transit chord of the planet to bypass any active latitudes on the stellar disc. The opposite is true when using starspot anomalies in light curves to determine $\lambda$. Due to this the two different methods complement each other in probing the dominant process in the dynamical evolution of transiting planets. It should be noted that in both the cases of WASP-19 and WASP-6 (see Fig.\,\ref{fig:con2}) the measured uncertainty in $\lambda$ is much smaller than measured using the RM effect. This indicates that the starspot method to measure $\lambda$ is superior to the RM effect in terms of reduced uncertainty in measuring $\lambda$. However, as was shown in observing WASP-50 (see \citealt{Jeremy2013}), the starspot method does not always work in terms of obtaining transit light curves affected by a starspot anomaly. The RM effect does have a high success rate in measuring a value of $\lambda$ but rarely achieves a similar precision.


\section{Acknowledgements}
\label{sec:Acknow}

We like to thank the anonymous referee for the helpful comments on the manuscript. The operation of the Danish 1.54\,m telescope at ESO’s La Silla observatory is financed by a grant to UGJ from The Danish Council for Independent Research (FNU). Research at the Armagh Observatory is funded by the Department of Culture, Arts \& Leisure (DCAL). JTR acknowledges financial support from STFC in the form of a Ph.D. Studentship (the majority of this work) and also acknowledges financial support from ORAU (Oak Ridge Associated Universities) and NASA in the form of a Post-Doctoral Programme (NPP) Fellowship. JS acknowledges financial support from STFC in the form of an Advanced Fellowship. DR acknowledges financial support from the Spanish Ministry of Economy and Competitiveness (MINECO) under the 2011 Severo Ochoa Program MINECO SEV-2011-0187. FF, DR (boursier FRIA) and J Surdej acknowledge support from the Communaut\'e fran\c{c}aise de Belgique -- Actions de recherche concert\'ees -- Acad\'emie Wallonie--Europe.



\end{document}